\documentclass[prb,reprint,amsmath,amssymb,aps,superscriptaddress]{revtex4-2}

\usepackage{amsmath,amsfonts,amssymb} %
\usepackage{graphicx}
\graphicspath{{figs/}}
\usepackage{dcolumn}
\usepackage{bm}
\usepackage[acronym]{glossaries}
\usepackage{tabularx} %
\usepackage{braket} %
\usepackage{multirow}
\usepackage[dvipsnames,table]{xcolor} %
\usepackage[breaklinks, %
colorlinks, linkcolor=Blue, citecolor=Blue, urlcolor=Blue]{hyperref} %
\usepackage{booktabs} %

\usepackage[ruled,vlined]{algorithm2e}
\usepackage{etoolbox}
\AfterEndEnvironment{strip}{\leavevmode}
\usepackage{graphicx}

\SetKwComment{Comment}{/* }{ */}
\SetKw{Require}{Require:}

\newcommand{\eq}[1]{Eq.~(\ref{#1})} %
\newcommand{\fig}[1]{Fig.~\ref{#1}} %
\def\be{\begin{equation}} %
\def\ee{\end{equation}} %
\newcommand{\bea}{\begin{eqnarray}}
\newcommand{\eea}{\end{eqnarray}}
\newcommand{\HO}{\hat O} %
\newcommand{\HP}{\hat P} %
\newcommand{\HU}{\hat U} %
\newcommand{\HH}{\hat H} %
\newcommand{\HZ}{\hat Z} %

\begin{document}
\setcitestyle{super}
    \title{Fidelity overhead for non-local measurements in variational quantum algorithms}

    \author{Zachary Pierce Bansingh}
    \affiliation{Chemical Physics Theory Group, Department of Chemistry, University of Toronto, Toronto, Ontario M5S 3H6, Canada}
      \affiliation{Department of Physical and Environmental Sciences, University of Toronto Scarborough, Toronto, Ontario M1C 1A4, Canada}
    \author{Tzu-Ching Yen}
    \affiliation{Chemical Physics Theory Group, Department of Chemistry, University of Toronto, Toronto, Ontario M5S 3H6, Canada}
    \author{Peter D. Johnson}
    \affiliation{Zapata Computing Inc., 100 Federal Street, Boston, Massachusetts 02110, USA}
    \author{Artur F. Izmaylov}
    \affiliation{Chemical Physics Theory Group, Department of Chemistry, University of Toronto, Toronto, Ontario M5S 3H6, Canada}
    \affiliation{Department of Physical and Environmental Sciences, University of Toronto Scarborough, Toronto, Ontario M1C 1A4, Canada}

    \date{\today}

    \begin{abstract}
    Measuring quantum observables by grouping terms that can be rotated to sums of only products of Pauli $\hat z$ operators 
    (Ising form) is proven to be efficient in near term quantum computing algorithms. This approach requires extra unitary transformations to rotate the state of interest so that the measurement of a fragment's Ising form would be equivalent to measurement of the fragment for the unrotated state. These extra rotations allow one to perform a fewer 
    number of measurements by grouping more terms into the measurable fragments with a lower overall estimator variance. 
    However, previous estimations of the number of measurements did not take into account non-unit fidelity of quantum gates  implementing the additional transformations. Through a circuit fidelity reduction, additional transformations introduce extra uncertainty and increase the needed number of measurements. Here we consider a simple model for errors introduced by additional gates needed in schemes involving grouping of commuting Pauli products. 
    For a set of molecular electronic Hamiltonians, we confirm that the numbers of measurements in schemes using non-local qubit rotations are still lower than those in their local qubit rotation counterparts, even after accounting for uncertainties introduced by additional gates.  
    \end{abstract}

    \maketitle
    \section{Introduction}
    
    Measuring quantum observables is one of the bottlenecks of quantum variational algorithms.\cite{gonthier2020identifying} 
    Here we consider quantum observables represented as a linear combination of Pauli products
  \bea
  \HO = \sum_k c_k \HP_k,
  \eea 
  where $c_k$ are coefficients and $\HP_k$ are tensor products of single-qubit Pauli operators.
  To measure an expectation value of $\HO$ one needs to 
  transform $\HO$ into the Ising form $\HU^\dagger \HO \HU = \hat Z$, where $\HZ$ is a linear combination of Pauli products that contain only $\hat z$ operators for individual qubits. Finding such $\HU$ for an arbitrary observable is challenging 
  since it is equivalent to diagonalizing $\HO$ in the computational basis. Fortunately, the expectation 
  value of $\HO$ can be obtained by summing the expectation values of simpler operators that 
  can be easily transformed to their Ising forms ($\HZ_n$): 
  \bea
  \bra{\Psi} \HO\ket{\Psi} = \sum_n \bra{\Psi} \HO_n\ket{\Psi} = \sum_n \bra{\Psi} \HU_n^\dagger \HZ_n \HU_n\ket{\Psi}.
\eea
   This is a partitioning approach to the measurement problem that was successfully applied in the variational quantum eigensolver (VQE),\cite{2014VQE} where $\HO$ corresponds to the system Hamiltonian, $\HH$. Even for a fixed observable 
 there are usually numerous possible partitionings $\HO = \sum_n \HO_n$, where each partitioning involves its own set of 
 unitary transformations $\HU_n$. For all partitionings $\HU_n$ should be straightforward to apply both on 
 classical and quantum computers because one needs to know in advance $\HZ_n$ operators for classical 
 post-processing and be able to apply $\HU_n$ to the quantum state $\ket{\Psi}$ before the measurement. 
 The main quantity of interest that ranks partitionings in their efficiency is the total number of measurements required for achieving accuracy $\epsilon$ in estimating $\bra{\Psi} \HO \ket{\Psi}$, $N_m$. This number 
 under the condition of optimal distribution of measurements between fragments $\HO_n$ is
 \bea\label{eq:nm}
        N_m=\left(\frac{1}{\epsilon}\sum_n\sqrt{\text{Var}_{\Psi}(\HO_n)} \right)^{2},
 \eea
 where $\text{Var}_{\Psi}(\HO_n) = \bra{\Psi} \HO_n^2 \ket{\Psi} -  \bra{\Psi} \HO_n \ket{\Psi}^2$ is the quantum 
 variance for each operator fragment.\cite{Crawford} 
 
 Since one of the possible areas for demonstrating the quantum advantage is the electronic structure problem, for 
 concreteness, here we focus on obtaining the expectation value of electronic Hamiltonians.    
 There are multiple ways one can partition the Hamiltonian into diagonalizable fragments. For electronic 
 Hamiltonians, there have been suggested two main approaches to partitioning using diagonalizable fragments in 
 fermionic and qubit operator algebras. Recent comparisons of state-of-the-art techniques in the two categories shown that 
 one can achieve a lower number of the measurements using techniques within the qubit operator algebra.\cite{CSA2021,Aadi_Overlapping,shlosberg2021adaptive} 
 These qubit techniques are based on partitioning of the Hamiltonian to sets of commuting 
 Pauli operators.\cite{yen2020measuring,MoscaA,gokhale2019minimizing,Verteletskyi:2020do,shlosberg2021adaptive} 
 
 Due to differences in accuracy of performing one- and two-qubit transformations in 
 quantum computers, it is convenient to distinguish two types of commutativity, general or full commutativity (FC) and 
 a more restrictive qubit-wise commutativity (QWC). Two Pauli products qubit-wise commute only if every pair of corresponding single-qubit operators commute in them (e.g. $\hat x_1 \hat y_2 \hat z_3$ qubit-wise commutes with $\hat x_1 \hat z_3$ and does not qubit-wise commute with $\hat y_1 \hat x_2 \hat z_3$). 
 Measuring a linear combination of Pauli products that qubit-wise commute require only local (one-qubit) 
 Clifford transformations ($\HU_n$), while for the same task on commuting Pauli products one generally needs non-local (entangling) Clifford transformations. The main advantage of grouping schemes based on FC over QWC schemes
  is lower numbers of measurements required for estimating the Hamiltonian expectation values (\eq{eq:nm}). 
  However, estimates based on \eq{eq:nm} do not account for additional uncertainties related to 
  non-unit fidelities of gates that are required to introduce extra transformations needed for measurement, $\HU_n$. 
  In this work, we investigate the extent to which these additional uncertainties reduce the advantage of the FC-based schemes over the QWC-based schemes in the number of measurements.

Classical shadow tomography (CST)\cite{Huang_2020} has been proposed recently as an alternative to grouping of operator terms in measurable fragments for estimating the operator expectation value.  CST techniques are based on learning about the expectation value of the operator by rotating the system with a random unitary transformations from a certain distribution and measuring the system in the computational basis. The obtained set of measurement results are projections (classical shadows) of the true system state. Expectation values of the operator for classical shadows can be calculated on a classical computer 
and averaged to reproduce the expectation value of the 
true system state. The Clifford group and its single-qubit subgroup distributions of unitaries have been used to draw 
random unitaries.\cite{Huang_2020,Zhao_2021,Hadfield_LBCS_2006,hadfield2021adaptive,PRXQuantum:POVM}  Using the electronic structure Hamiltonians as examples of quantum observables, it was shown that for a single observable, CST schemes are generally inferior than grouping techniques in lowering the number of measurements required to estimate the expectation value.\cite{Aadi_Overlapping,shlosberg2021adaptive} 
Therefore, here we consider the grouping methods to measurements but 
our analysis involving Clifford transformations can be extended to the CST schemes as well.

     \section{Method}

\subsection{Error Model}

In both FC and QWC partitionings the electronic Hamiltonian in the qubit representation is 
partitioned to fragment Hamiltonians, $\hat H = \sum_{n=1}^{N_f}\hat H_n$, 
    \begin{equation}
         \hat H_n = \sum_{j=1}^{M_n} a_{j}P_{j},
    \end{equation}
    where $P_{j}$ are Pauli products, and $P_{j}$ within each $\hat H_n$ commute or qubit-wise commute with each other. 
    For all $\hat H_n$, there are unitary transformations from the Clifford group $\hat U_n$ so that 
    $\hat H_n = \hat U_n^\dagger \hat Z_n \hat U_n$. 

The VQE workflow can be separated in two parts: 1) preparing the state $\ket{\Psi} = \hat U \ket{0}$ 
and 2) transforming $\ket{\Psi}$ to the state that will be measured: $\ket{\Phi_n} = \hat U_n \ket{\Psi}$. 
The total fidelity of the $\ket{\Phi_n}$ state preparation $F_n = p q_n$ is a product of fidelities for the preparation 
$\ket{\Psi}$, $p$, and of that for the additional measurement preparatory circuit $\hat U_n$, $q_n$. 
FC and QWC schemes will be different in values of $q_n$ and share the $p$-part. 
Since the total fidelity is multiplicative we assume that the common $p$-part 
also incorporates any noise contributions appearing due to projective measurements of all qubits. For the $q_n$-part, 
fidelities of one- and two-qubit gates are denoted as $f_1$ and $f_2$, respectively. Thus, if $G_{i,n}$ are the numbers of 
$i$-qubit gates for the $\hat U_n$ circuit then $q_n = f_1^{G_{1,n}} f_2^{G_{2,n}}$.  
    
%
   
    Due to non-unit gate fidelities in the state preparation we assume that instead of $\ket{\Phi_n}$ 
    one can only prepare the mixed state    
    \begin{equation}\label{eq:rhon}
        \rho_n = F_n \ket{\Phi_n}\bra{\Phi_n} +  \frac{(1-F_n)}{d}\mathbf{1},
    \end{equation}
    where $d=2^{N_q}$ is the dimension of the $N_q$-qubit space, and $\mathbf{1}$ is the identity operator in the $d$-dimensional space.  In \eq{eq:rhon}, it is assumed that there is no systematic bias and extra gates only contribute to probability of the depolarized component. 
   
    \subsection{Estimator characteristics}
    
    Here we construct an estimator for $\bra{\Psi} \hat H_n \ket{\Psi}$ assuming that we only have access to measurement 
    results for the density $\rho_n$ (\eq{eq:rhon}) on the transformed fragment Hamiltonian, $\hat H_n = \hat U_n^\dagger \hat Z_n \hat U_n$
    \bea
        {\rm Tr}[\hat Z_n \rho_n] &=& F_n \bra{\Phi_n} \hat Z_n \ket{\Phi_n} + (1-F_n)\text{Tr}[\hat Z_n]/d \\
        &=& F_n \bra{\Phi_n} \hat Z_n \ket{\Phi_n} = F_n \bra{\Psi} \hat H_n \ket{\Psi},
    \eea
    where the second equality follows from the tracelessness of $\hat Z_n$.
    Therefore, an estimator for $\bra{\Psi} \hat H_n \ket{\Psi}$, $\bar{H}_n$, can be formulated as  
    \bea
        \bar{H}_n = \frac{{\rm Tr}[\hat Z_n \rho_n]}{F_n}.
    \eea
    Its statistical variance can be obtained from the quantum variance of ${\rm Tr}[\hat Z_n \rho_n]$
    \begin{equation}
        \text{Var}(\bar{H}_n) = \frac{{\rm Tr}[\hat Z_n^2 \rho_n]-{\rm Tr}[\hat Z_n \rho_n]^2}{F_n^2}.
    \end{equation}
    Next, we would like to connect the statistical variance of our estimator with the quantum variance of the fragment 
    $\text{Var}_{\Psi}(\hat H_n) = \bra{\Psi} \hat H_n^2 \ket{\Psi}- \bra{\Psi} \hat H_n \ket{\Psi}^2$. 
    This connection will elucidate a functional dependence of the estimator variance on 
    fidelity contributions and the operator properties. Using $\rho_n$ form in \eq{eq:rhon} gives 
    \bea \notag
    \text{Var}(\bar{H}_n) &=& \frac{\bra{\Psi} \hat H_n^2 \ket{\Psi}}{F_n} +\frac{1-F_n}{dF_n^2} {\rm Tr}[\hat Z_n^2] \\
    &&-\bra{\Psi} \hat H_n \ket{\Psi}^2 \\ \notag
    &=& \text{Var}_{\Psi}(\hat H_n) + \frac{1-F_n}{F_n} \bra{\Psi} \hat H_n^2 \ket{\Psi} \\
    &&+ \frac{1-F_n}{dF_n^2} {\rm Tr}[\hat Z_n^2].\label{eq:HvarP}
    \eea
    Both terms additional to $\text{Var}_{\Psi}(\hat H_n)$ in \eq{eq:HvarP} are positive. ${\rm Tr}[\hat Z_n^2]/d$  can be further 
    simplified as ${\rm Tr}[\hat Z_n^2]/d = \sum_{j=1}^{M_n} a_j^2$ since any product of two different Pauli products is traceless. 
    This gives the final expression for the estimator variance as
    \bea\notag
     \text{Var}(\bar{H}_n) &=& \text{Var}_{\Psi}(\hat H_n) + \frac{1-F_n}{F_n} \bra{\Psi} \hat H_n^2 \ket{\Psi} \\
     &&+ \frac{1-F_n}{F_n^2} \sum_{j=1}^{M_n} a_j^2 .\label{eq:Hvar}
    \eea
 This expression allows us to formulate the minimal number of measurements for achieving $\epsilon$ accuracy with 
 $66.7\%$ probability as 
     \begin{equation}\label{eq:bb2}
        N_m=\left(\frac{1}{\epsilon}\sum_n\sqrt{\text{Var}(\bar{H}_n)} \right)^{2}.
    \end{equation}
Due to the relation $\text{Var}(\bar{H}_n) > \text{Var}_{\Psi}(\hat H_n)$ for $p \ne 1$ or $q_n \ne 1$, $N_m$ will always be 
larger than its counterpart evaluated ignoring non-unit fidelity of gates. 

To understand  why lower fidelities can be more detrimental for the QWC partitioning than for the FC one, 
it is instructive to rewrite $\text{Var}(\bar{H}_n)$ as
    \bea\notag
     \text{Var}(\bar{H}_n) &=& \text{Var}_{\Psi}(\hat H_n) + \frac{1-F_n}{F_n} \sum_{i\ne j} a_ia_j\bra{\Psi} \hat P_i\hat P_j \ket{\Psi} \\
     &&+(1-F_n)(F_n^{-1}+F_n^{-2}) \sum_{j=1}^{M_n} a_j^2, \label{eq:Hvar2}
    \eea
    where we partitioned the $\bra{\Psi} \hat H_n^2 \ket{\Psi}$ term into the diagonal ($i=j$) and off-diagonal ($i\ne j$) parts.
    For lower fidelities the last term in \eq{eq:Hvar2} will dominate because of the $F_n^{-2}$ scaling. 
    In the case where $F_n$ is dominated by the uniform reduction factor $p$, the estimator variance for each 
    fragment can be approximated as
         \bea
     \text{Var}(\bar{H}_n) &\approx& \frac{(1-p)}{p^2} \sum_{j=1}^{M_n} a_j^2, \label{eq:HvarA}
    \eea
and the total number of measurements as
         \bea\label{eq:peff}
         N_m \approx N_m^{(p)} = \frac{(1-p)}{\epsilon^2 p^2}\left(\sum_n\sqrt{\sum_{j=1}^{M_n} a_j^2} \right)^{2}.
    \eea
Clifford group transformations do not change coefficients $a_j$, hence the total sum over all $a_j^2$ in all 
groups is a constant dependent on the system Hamiltonian rather than the partitioning approach. 
Clearly, this constant is partitioned to smaller parts in the QWC grouping than in the FC one. It is a general 
property of a square root function that $\sqrt{a+b}\le \sqrt{a}+\sqrt{b}$ for real positive $a$ and $b$. Thus, if the 
total sum of $a_j^2$ is split into a larger number of smaller groups in QWC, 
and QWC $N_m^{(p)}$ will be larger than that for the FC scheme. 
This also follows from the fact that any QWC group is also FC group but not the other way around.     
    
    \section{Results}
\begin{figure*}[!ht]
    \center
    \includegraphics[width=1\textwidth]{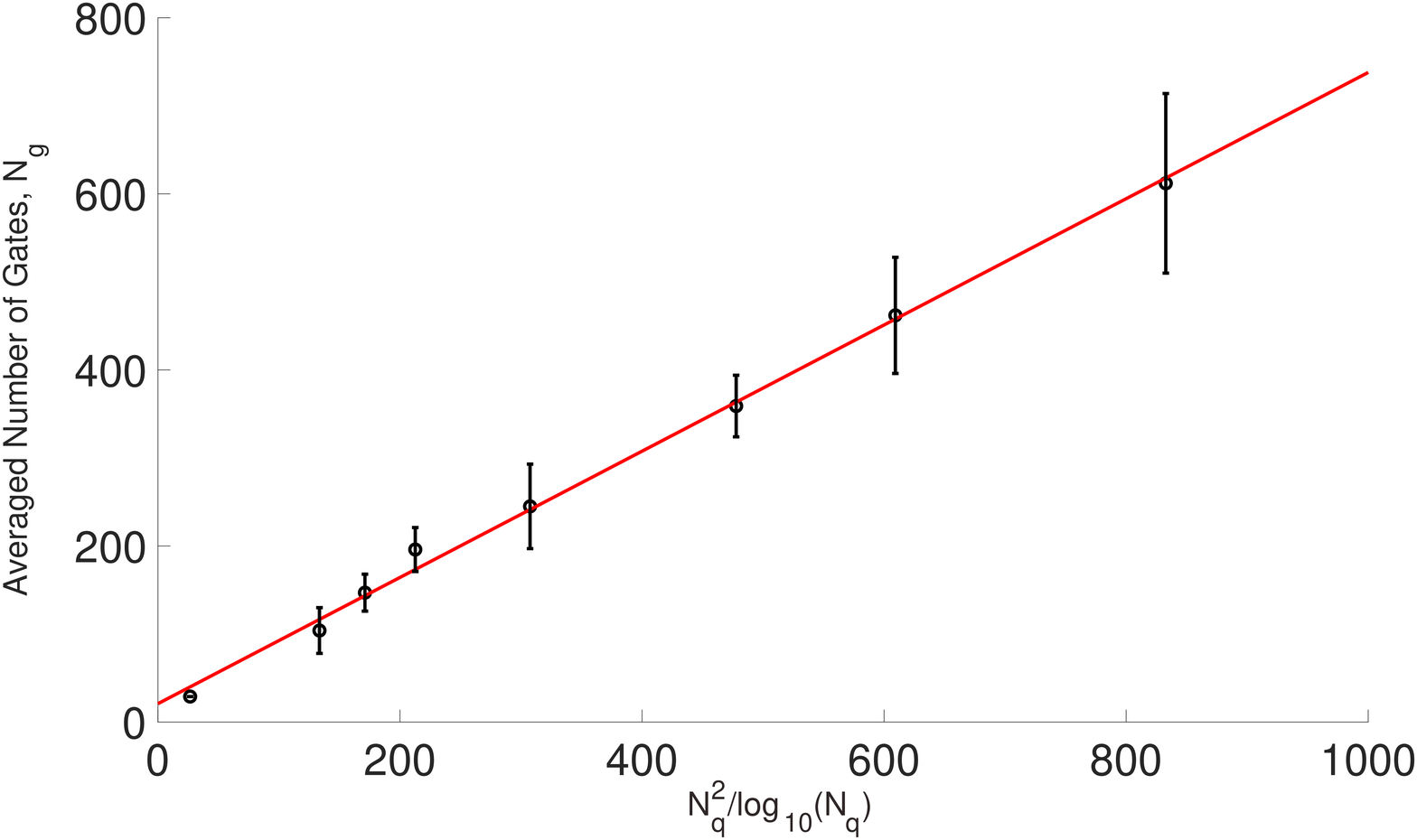}
  \caption{\label{Gate Complexity} An average number of gates ($N_g$) for measurement circuits of the FC scheme as a function of $N_q^2/\log_{10}(N_q)$, where $N_q$ is the number of qubits. The averaging is done over measurable groups of the FC scheme for molecular systems described in Appendix A. The linear function $N_g = 0.72 N_q^2/\log_{10}(N_q) + 21$ (red) has $R^2 = 0.997$.}
      \label{fig:scaling}
  \end{figure*}

    We assess the difference in the number of measurements required in the FC and QWC schemes on a set of molecular electronic Hamiltonians. Details on the Hamiltonians are presented in Appendix \ref{sec:hamdet}. Among various heuristics for grouping commuting Pauli products\cite{Verteletskyi:2020do,yen2020measuring,tilly2021variational} we used the Sorted Insertion (SI) method\cite{Crawford} because it generally outperforms other grouping techniques in the number of measurements.\cite{CSA2021,Aadi_Overlapping} 
Since measurement circuits for $\HU_i$ of the FC and QWC groups can be constructed 
 using only Clifford group transformations, according to the Gottesman-Knill theorem 
 these circuits can be implemented efficiently on both classical and quantum computers.\cite{gottesman1998heisenberg,10.5555/1972505}  
 For the FC grouping an asymptotically optimal scaling of $O(N_q^2/\log(N_q))$ in the number of 2-qubit entangling gates and single qubit rotations can be achieved.\cite{ImprovedSimulation,patel2003efficient} We have implemented this optimal circuit synthesis and optimization method in the Tequila program\cite{2021Tequila}. The total number of CNOT, Hadamard and Phase gates for model systems averaged over measurable FC groups indeed follows 
 the $O(N_q^2/\log(N_q))$ dependence according to \fig{fig:scaling}. 
    
   Considering variety in fidelities of quantum gates available in different 
    near term quantum computing architectures, we calculated $q_n$-parts of the total fidelity using two sets of 1- and 2-qubit gate fidelities. One set is based on the numbers from a recent work describing the Google superconducting quantum hardware,\cite{arute2019quantum} $f_1=0.9938$ and $f_2 = 0.9984$. This set represents what is available on a contemporary superconducting qubit device. The second set is based on a projection that in the near future we will have quantum gates with $f_1 = 0.9999$ and $f_2 = 0.999$. The $p$-component of the total fidelity depends
    not only on gate fidelities but also on the state complexity, here, we would like to see the trend by evaluating the number of 
    measurement estimates for $p=1$, $p=0.8$, and $p=0.6$.  
    
    Table \ref{tab:FCvsQWC} provides the ratios of the measurement numbers required in the FC and QWC schemes.  Ideal FC (FC-I) has always lower $N_m$ than that of QWC (QWC-I) except for H$_2$. The ratio between the numbers of measurements  grows with the system size and reaches more than 20 for the largest system (N$_2$). 
    When we introduce uncertainty related to non-unit fidelity of gates, the ratios decrease but in most cases indicate 
    superiority of the FC scheme. This reduction in ratios is the consequence of generally lower fidelities 
    ($F_n$) in FC groups compared to those of QWC groups because of a larger number of gates needed for 
    measurements in the former. For small systems, H$_2$ and LiH, non-unit fidelity of gates lead to lower numbers of measurements for the QWC scheme (ratios $<1$). Interestingly, accounting for the reduction of the circuit fidelity 
    due to the $\ket{\Psi}$ state preparation ($p=0.8$ and $p=0.6$) increases the ratios for the small systems and even 
    reverses the trend for H$_2$ and for LiH with the second set of gate fidelities. This detrimental effect of $p$ reduction on 
    QWC $N_m$'s is attributed to the contribution of \eq{eq:peff} that favours a smaller number of groups. The effect from  
    \eq{eq:peff} generally competes with other terms in \eq{eq:Hvar2} and is clearly seen in cases 
    of H$_2$, LiH, and NH$_3$ (gate fidelities of a contemporary device). For the other systems, $p$ reduction leads to reduction     
    of the ratios.

       \begin{table*}[!htbp]
        \setlength\tabcolsep{0pt}
        \caption{Ratios of the number of measurements $N_m$ to achieve $\epsilon=1 \times 10^{-3}$ Hartree accuracy in the energy expectation value (with 66.7\% probability) assuming the optimal distribution of measurements for individual fragments. 
        Two partitioning methods are considered: FC and QWC. Three gate fidelity sets are used: ideal, unit fidelities (-I); fidelities available on a contemporary device,   set 1 (-C); and near future fidelities, set 2 (-F). } 
            \centering
        {\begin{tabular*}{1\textwidth}{@{\extracolsep{\fill}}  l l l l l l l l}
        \toprule
         Systems & H$_2$ & LiH & BeH$_2$ & H$_2$O & NH$_3$ & N$_2$ \\
        $N$ & 4 & 12 & 14 & 14 & 16 & 20 \\
        \midrule
     \multicolumn{7}{c}{$p=1$} \\
          $N_m$(QWC-C)/ $N_m$(FC-C) & 0.965 & 0.602 & 1.91 & 1.15 & 1.20 & 1.67 \\
         $N_m$(QWC-F)/ $N_m$(FC-F)  & 0.993 & 0.990 & 4.54 & 2.72 & 4.17 & 14.1 \\ 
         $N_m$(QWC-I)/ $N_m$(FC-I) & 1.00 & 1.08 & 5.51 & 3.24 & 5.30 & 23.2 \\ 
        \midrule
        \multicolumn{7}{c}{$p=0.8$} \\
         $N_m$(QWC-C)/ $N_m$(FC-C) & 1.04 & 0.912 & 1.37 & 1.14 & 1.18 & 1.24 \\
         $N_m$(QWC-F)/ $N_m$(FC-F) & 1.06 & 1.11 & 1.68 & 1.36 & 1.75 & 2.12 \\ \midrule
          \multicolumn{7}{c}{$p=0.6$} \\
         $N_m$(QWC-C)/ $N_m$(FC-C) & 1.09 & 0.979 & 1.29 & 1.15 & 1.20 & 1.20 \\
         $N_m$(QWC-F)/ $N_m$(FC-F) & 1.10 & 1.16 & 1.55 & 1.33 & 1.65 & 1.85 \\
        \bottomrule
        \end{tabular*} 
        }
        \label{tab:FCvsQWC}
    \end{table*}
  
    \section{Conclusions}
    
Using a simple unbiased depolarization error model we have derived an estimation 
for the number of measurements needed in the VQE measurement schemes involving local and non-local qubit transformations. 
Considering realistic fidelities for current and near future superconducting quantum processors and using the implementation of the non-local unitary transformations minimizing the number of CNOT gates, it was found that the non-local scheme based on grouping commuting Pauli products generally requires  fewer measurements than the corresponding local scheme based on grouping of qubit-wise commuting Pauli products. 
Due to a larger number of additional gates the advantage of the non-local scheme decreases with reduction in gate fidelities. Another trend is that the advantage from the FC scheme is diminished even further if one accounts for the fidelity reduction 
due to the state preparation circuit. However, lowering of the state preparation fidelity can also exhibit an interesting opposite trend that favors the FC scheme because it has a smaller number of groups. 
Overall, we found that for all systems larger than LiH (in the STO-3G basis) the number of measurements 
needed in the FC scheme is lower than that in the QWC scheme even with accounting for 
all uncertainties introduced by non-unit gate fidelities.

    \section*{Acknowledgements}
     A.F.I. acknowledges financial support from the Google Quantum Research Program, Early Researcher Award, and Zapata Computing Inc. This research was enabled in part by support provided by Compute Ontario and Compute Canada.

     \appendix
   
    \section{Details of Hamiltonians}\label{sec:hamdet}

    Qubit Hamiltonians were generated using the STO-3G basis and the Bravyi-Kitaev (BK) transformation.\cite{Bravyi:2002/aph/210} The nuclear geometries for the Hamiltonians are 
    R$(\rm H-H) = 1 \AA$  ($\rm H_2$), R$(\rm Li-H) = 1 \AA$ ($\rm LiH$), R$(\rm Be-H) = 1 \AA$ with collinear atomic arrangement ($\rm BeH_2$), 
    R$(\rm O-H) = 1 \AA$ with $\angle HOH = 107.6^\circ$ ($\rm H_2O$), R$(\rm N-H) = 1 \AA$ with $\angle HNH = 107^\circ$ ($\rm NH_3$), and R$(\rm N-N) = \AA$ ($\rm N_2$). To collect more data points for circuit gate complexity in \fig{fig:scaling}, additional qubit Hamiltonians were generated using the 6-31G basis and the BK transformation 
    for $\rm BeH_2$, $\rm H_2O$, $\rm NH_3$, and $\rm N_2$ with the same nuclear geometries as in the STO-3G case.

\end{document}